\documentclass[twocolumn,showpacs,preprintnumbers,amsmath,amssymb]{revtex4}

\usepackage{graphicx}
\usepackage{dcolumn}
\usepackage{bm}

\begin{document}


\title{Spin Wave Propagation in the Domain State of a Random Field Magnet}

\author{R. L. Leheny$^{1}$, 
Y. S. Lee$^{2,3}$, G. Shirane$^{4}$, and R. J. Birgeneau$^{3,5}$}

\affiliation{$^{1}$Department of Physics and Astronomy, Johns Hopkins
University,
Baltimore, MD 21218, USA
\\
$^{2}$NIST Center for Neutron Research, National Institute of 
Standards and Technology, Gaithersburg, MD 20899
\\
$^{3}$Department of Physics, 
Massachusetts Institute of Technology,
Cambridge, MA 02139, USA
\\
$^{4}$Department of Physics, Brookhaven National Laboratory, Upton, NY 
11973
\\
$^{5}$Department of Physics, University of Toronto, Toronto, Canada 
M5S 1A1
}

\date{\today}

\begin{abstract}
Inelastic neutron scattering with high wave-vector 
resolution has characterized the propagation of transverse spin wave 
modes near the 
antiferromagnetic zone center in the metastable domain state 
of a random field Ising magnet.  A well-defined, long wavelength 
excitation is observed despite the absence of long-range magnetic order.
Direct comparisons with the spin wave dispersion in the 
long-range ordered antiferromagnetic state reveal no measurable effects from the domain 
structure.  This result recalls analogous behavior in thermally 
disordered anisotropic spin chains but contrasts sharply with that of
the phonon modes in relaxor ferroelectrics.  

\end{abstract}
\


\maketitle

Disorder in a condensed matter system can strongly influence mode 
propagation through the medium, providing a unique perspective on its 
microscopic behavior.  To gain a more comprehensive understanding of 
such effects from disorder, we have performed a neutron scattering 
study of the long wavelength spin waves of the frozen domain state of 
a random field magnet.  Despite the wealth of attention directed at the random field 
problem~\cite{birgeneau-review}, little attention has been directed 
toward the spin wave dispersion 
in these systems.  Our study has been motivated in part by 
experiments on the phonons in 
several relaxor ferroelectrics including Pb(Zn$_{1/3}$Nb$_{2/3}$)O$_{3}$ 
and Pb(Mg$_{1/3}$Nb$_{2/3}$)O$_{3}$ that have demonstrated interesting 
anomalies associated with the disorder in these 
systems~\cite{gehring-prb,gehring-preprint,gehring-prl,koo}.  
Specifically, at wave vectors near 0.2 \AA$^{-1}$, the
dispersion of the transverse optic branch appears to drop precipitously into the acoustic 
branch.  This phenomenon has been identified with a sharply 
wave-vector dependent overdamping 
associated with the presence of polar nanoregions
in the relaxors~\cite{gehring-prb}.
To test the generality of such behavior for systems with disorder, 
we have performed a search for analogous effects in the spin 
waves of the domain state of the random field 
magnet.  In marked contrast to the phonons
in the relaxors, we observe well-defined long wavelength excitations 
transverse to the ordering axis, despite the 
absence of long range order, and find no measurable effect on the spin 
waves from the domain structure.  We identify the 
persistence of these modes as an apparently generic feature of disordered magnetic 
systems with anisotropy, as 
first realized in thermally disordered spin 
chains~\cite{villain,CsNiF3,TMMC}.  
 
The experiments were performed on a diluted 
antiferromagnet, Mn$_{x}$Zn$_{1-x}$F$_{2}$ with $x=0.5$, cooled in a 
magnetic field parallel to the Ising axis.   
The physics of the diluted Ising antiferromagnet in a uniform field can 
be mapped directly on to that of a random field Ising 
magnet~\cite{aharony}, providing a method for experimental realization 
of the random field system.
MnF$_2$ has a tetragonal rutile-type 
structure with lattice constants $a=4.87$ {\AA} and $c=3.31$ {\AA}.
The spin interaction is primarily Heisenberg-like, and a weak dipolar interaction 
between the Mn moments accounts for the small Ising anisotropy that 
aligns the Mn spins along the c axis (i.e., the tetragonal axis).
Diluted MnF$_2$ was 
chosen for this study because, as a weakly Ising antiferromagnet, its small spin gap
makes the physics of interest easily accessible.  
We stress that, while the interesting phonon behavior in relaxors 
motivates our spin wave study on this system, the diluted antiferromagnet in field
should not be considered a precise magnetic analog of disordered 
ferroelectrics.  First, the antiferromagnet does not share the 
property of a conserved order parameter.  In addition, the nature of 
the polar nanoregions in the relaxors as the source of random fields 
has recently come into question~\cite{xu}.   
Never-the-less, in pursuing the general 
problem of the effects of disorder on the spin waves in a 
random field magnet, we view diluted MnF$_2$ as the best 
candidate for experiments.

When cooled from 
the paramagnetic phase in the field, the diluted antiferromagnet forms a 
metastable state of frozen domains.  Using high resolution neutron 
scattering, we have characterized this domain structure
and the corresponding spin wave dispersion at wavelengths larger than the typical
domain size.  In addition, by cooling the antiferromagnet in 
zero field and then applying the field, we have produced a long-range ordered 
antiferromagnetic state in the system.  Comparing the results obtained 
after cooling in field with the results of identical measurements taken after zero 
field cooling, we are able to determine directly the influence of the 
domain structure on the spin waves.

An important criterion for this experiment is the ability to create 
domains under field cooling that are sufficiently small to resolve 
the spin wave dispersion at the appropriate wave vectors.  Smaller domains are 
achieved with higher dilution~\cite{birgeneau,cowley}, 
larger applied fields~\cite{cowley}, and faster cooling 
rates.  With too high dilution, however, the spin waves (in zero field) 
eventually become poorly defined~\cite{uemera}.  In addition, the presence of 
a spin-flop transition in Mn$_{x}$Zn$_{1-x}$F$_{2}$, with a 
spin-flop field that depends on $x$~\cite{cowley}, limits the maximum field 
that
one can apply while remaining in the random field Ising regime.   Previous 
studies~\cite{birgeneau,cowley,uemera} indicate that $x=0.5$ is the 
optimal compromise among these different concerns.   The experiment 
was performed at the NIST Center for Neutron Research 
on the SPINS triple-axis spectrometer, with tight collimation 
(30'-10'-S-10'-20') to achieve high wave-vector resolution
and a final neutron energy fixed at 3.5 meV\@.  A single crystal of 
Mn$_{0.5}$Zn$_{0.5}$F$_{2}$, approximately 50 grams in size, was 
oriented with the c-axis 
(the Ising axis) parallel to the external field and perpendicular to 
the scattering plane.   The crystal was strongly coupled to the bath of a helium 
flow cryostat, permitting rapid cooling ($>20$ K/minute) from the high 
temperature paramagnetic phase to 4.2 K, well below the N\'{e}el 
temperature in zero field, $T_{N}(H=0)=21$ K.

Figure 1 displays scattering profiles from elastic scans through the 
antiferromagnetic zone center (1, 0, 0) along the transverse direction 
$(1, q_{K}, 0)$ in an 
external field of $H=2.3$ T.   The open circles 
represent measurements taken after cooling the sample in zero field, 
so that the random field strength is zero through the Ising transition, 
and then raising the field (ZFC).  The scattering 
follows a resolution-limited lineshape consistent with long-range 
antiferromagnetic order.  The solid circles show the results from cooling in field (FC) 
and display a significantly broadened lineshape corresponding to the 
metastable domain state induced by the 
random fields.  The scattering in the field-cooled 
measurement also shows a large overall enhancement in intensity, which we associate with 
the relief of extinction due to the domain structure.  We model the 
short-ranged correlations of this field-cooled domain state with a 
Lorentzian-squared form,

\begin{equation}
    I({\bf q},\omega=0)  = 
\frac{\sigma}{(1+{\bf q}^{2}\xi^{2})^{2}}
\end{equation} 

\noindent
where ${\bf q}$ is the wave vector measured with respect to the 
antiferromagnetic zone center.  The solid line in Fig.~1 is the 
results of a fit with Eq.~(1) convolved with the instrumental 
resolution function.  This static neutron cross section is proportional to the Fourier transform of the real-space two spin correlation function.  The transform of Eq.~(1) corresponds 
to an exponential decay in the correlations between Ising spins,

\begin{equation}
    \langle S_{z}(r)S_{z}(0) \rangle \sim e^{-r/\xi}.
\end{equation} 

\noindent
The Lorentzian-squared form for the elastic cross section has been shown in previous studies of field-cooled Mn$_{x}$Zn$_{1-x}$F$_{2}$~\cite{birgeneau,cowley} and
other random field Ising magnets, including Co$_{x}$Zn$_{1-x}$F$_{2}$~\cite{hagen} and 
Rb$_{2}$Co$_{x}$Mg$_{1-x}$F$_{4}$~\cite{birgeneau83}, to describe accurately the static
short-range order in the field-cooled state.  
In addition, theoretical arguments have indicated that this form 
should be expected on general 
grounds~\cite{lovesey,joanny,vilfan,nattermann}. 
Thus, field-cooled state can be considered a collection of 
metastable domains whose size distribution produces the
real-space correlation function given by Eq.~(2).
The correlation length 
extracted from the fit with Eq.~(1), $\xi = 137 \pm 4$ \AA, sets the characteristic 
length scale 
in the a-b plane for the domains.  Assuming that the domain structure
is isotropic in units of the lattice spacings, as 
suggested by the crystal structure, the smaller lattice constant along 
the c-axis leads to a characteristic scale of $93 \pm 3$ {\AA} in that 
direction.

\begin{figure}
 \centering\includegraphics[scale=0.5]{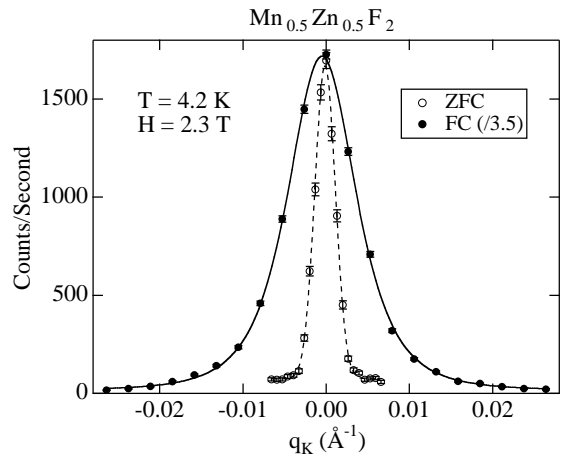}
\caption{Neutron scattering intensity at zero energy transfer 
scanning through the antiferromagnetic zone center (1, 0, 0) along the 
direction (0, $q_{K}$, 0) at $T=4.2$ K and $H=2.3$ T.  The open circles are the results after zero 
field cooling and have a resolution-limited lineshape.  The solid 
circles are the results after field cooling and have a lineshape 
reflecting the short-range ordered domain state.  The field cooled 
intensities have been divided by 3.5 for comparison.  The solid line is a 
fit to Eq.~(1) convolved with the instrumental resolution.}
\label{fig1}
\end{figure}

To compare the spin wave behaviors in the long-range ordered and domain 
states, we have measured the scattering intensity as a function of energy
transfer, $\omega$, at $q=0$ through the dispersion, as shown in Fig.~2.  Because 
of the finite experimental resolution, this measurement effectively 
integrates over the approximate wavevector ranges $-0.004 
$\AA$^{-1}< q_{H} <0.004 $\AA$^{-1}$, $-0.003 $\AA$^{-1}< q_{K} <0.003 
$\AA$^{-1}$ within the scattering plane and 
$-0.085 $\AA$^{-1}< q_{L} <0.085 $\AA$^{-1}$ perpendicular to 
it~\cite{collimation1,collimation2}.  
Based on the phonon behavior observed in the relaxor ferroelectrics, we
expect any change caused by the domain structure to affect the spin waves at 
wave vectors less than $q \approx 2\pi/\xi = 0.046$ \AA$^{-1}$ along H 
and K and $q \approx 2\pi/\xi = 0.068$ \AA$^{-1}$ along L\@.  Thus, 
a redistribution of spectral weight to lower 
energy, like the dramatic overdamping seen in the relaxors,
should be clearly visible in the scattering intensity in Fig.~2.  
As the figure illustrates, the results for the long-range ordered antiferromagnet and 
domain state are essentially identical through the dispersion at 
$\omega=0.5$ meV, indicating that the long wavelength spin waves are 
insensitive to the domain structure.  The scattering at smaller energy transfers, shown in 
the inset to Fig.~2, does display enhanced intensity in the domain 
state.  However, we associate this enhancement with the stronger 
Bragg scattering observed for the domain state in Fig.~1, extending 
away from $\omega=0$ due to the finite experimental resolution, and 
not to any change in the spin wave behavior.  In particular, any 
redistribution of spectral weight should be reflected in both enhanced 
scattering at small $\omega$ and reduced scattering at larger 
$\omega$, which is excluded by the results in Fig.~2.  This
dramatic contrast with the phonons in relaxors lends additional support  
to recent evidence questioning the accuracy of a random field picture 
for the relaxors~\cite{xu}.
 
\begin{figure}
\centering\includegraphics[scale=0.45]{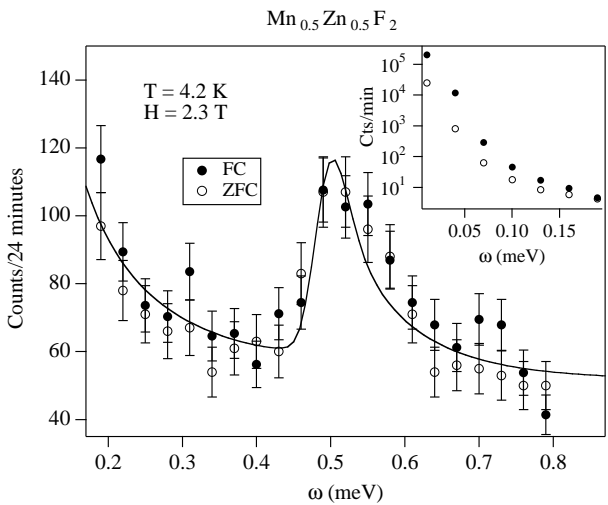}
\caption{Neutron scattering intensity as a function of energy 
transfer, $\omega$, at the antiferromagnetic zone center (1, 0, 0)
after field cooling (solid circles) and 
zero field cooling (open circles).  The essentially identical 
profiles through the spin wave dispersion at $\omega \approx 0.52$ meV 
demonstrates that the long wavelength spin waves are insensitive to 
the domain structure.  The solid line is the calculated result for the spin wave 
peak plus background based on the spectrometer resolution function
and the known dispersion relation in 
Mn$_{0.5}$Zn$_{0.5}$F$_{2}$ [13], assuming the that 
intrinsic width of the dispersion is zero.  
The enhanced intensity after field cooling at 
small $\omega$, shown in the inset, reflects the enhanced Bragg 
scattering due to relief of extinction.}
\label{fig2}
\end{figure}
 
The solid line in Fig.~2 is the calculated result for the spin wave 
peak plus background based on the spectrometer resolution function
and the previously measured dispersion relation in 
Mn$_{0.5}$Zn$_{0.5}$F$_{2}$~\cite{uemera}, assuming the that 
intrinsic width of the dispersion is zero.  The only free 
parameters entering the calculation are the overall amplitude of the 
peak and the form of the sloping background, $A\omega^{-2}+B$.
The strongly asymmetric lineshape for the calculated result derives from 
the non-zero slope of the dispersion relation away from the zone center and the relatively 
coarse wave-vector resolution perpendicular to the scattering 
plane~\cite{collimation2}.  Any small discrepancies between the calculation and measurements 
likely result from uncertainties in 
this contribution to the resolution and in the dispersion relation, 
implying that the long wavelength spin waves in both field cooled 
and N\'{e}el states are long-lived.  Furthermore, direct comparisons 
of the amplitudes and widths of the measured FC and ZFC peaks 
indicate that they are the 
same to within 5\%, setting a small upper bound on any possible broadening 
specific to the field cooled domain state.   

This clear insensitivity of the long wavelength spin waves to the static disorder of the FC domain structure raises two interesting issues that merit further study.  The first concerns the effect on short wavelength spin waves.  A comparison between the spin wave dispersions in the FC and ZFC states at wave vectors close the antiferromagnetic zone edge would address this question.  As shown in Ref.~\cite{uemera}, the spin wave dispersion at wave vectors away from the zone center in a diluted Ising antiferromagnet becomes increasingly diffuse due to the geometric disorder of the dilution.  While this effect could severely complicate comparisons between FC and ZFC behaviors, it also indicates that short wave length spin waves are strongly affected by disorder.  The second issue for further study concerns the temperature dependence of the spin wave dispersion, particularly in the transition region.  The interplay of the critical behavior and the disorder-induced slow dynamics in the random field Ising magnet has presented serious challenges to a comprehensive description of the transition~\cite{birgeneau95, hill, belanger,ferreira}.  A characterization of the evolution with temperature of the spin wave dispersion near the zone center on approach to the transition would provide a new perspective on this issue.  

The energy gap of approximately 0.52 meV for the excitation in Fig.~2 matches that expected 
from the Heisenberg exchange field and dipolar 
anisotropy of Mn$_{0.5}$Zn$_{0.5}$F$_{2}$, approximately 
0.27 meV~\cite{uemera}, and the Zeeman 
energy from the precession around the applied field, 0.25 meV for a 
2.3 T field.  This excitation, 
the gapped mode of the Ising state, corresponds to long wavelength oscillations in 
the spin components transverse to the easy axis.  Because of the 
experimental resolution, the peak in Fig.~2 represents contributions 
from oscillations over a range of wavelengths, most of which far 
exceed the correlation length for magnetic order in the domain state.  
The existence of these modes, despite the disorder, bears a striking 
similarity to the behavior in thermally 
disordered spin chains with easy plane anisotropy.  As first predicted 
by Villain~\cite{villain}, 
well-defined long wavelength transverse modes have been observed in a number 
of one-dimensional anisotropic spin systems including CsNiF$_{3}$~\cite{CsNiF3} and 
(CD$_{3}$)$_{4}$NMnCl$_{3}$~\cite{TMMC}.  The measurement of such 
excitations in the domain state of a three-dimensional random field 
Ising magnet indicates that these long-lived transverse modes occur more generally 
than previously realized,
irrespective of the nature of the anisotropy, the system's spatial 
dimension, or the type of disorder.  Indeed, studies of spin 
glasses appear to support this picture that these modes are generic to 
anisotropic disordered magnets.  
While spin waves in canonical Heisenberg spin 
glasses~\cite{binder&young} are strongly overdamped, 
well-defined long wavelength 
excitations have been observed in Ising spin glasses~\cite{wong}.  
Theoretical insight into the apparently universal nature of these 
transverse modes in disordered 
anisotropic magnets would be 
invaluable.  

We thank Y. Chen for assistance with the data analysis and C. Broholm 
and P. Gehring for helpful discussions.  Financial support by the U. 
S. Dept.~of Energy under Contract No.~DE-AC02-98CH10886 and by ONR 
under project MURI (N00014-96-1-1173) is acknowledged.  Research
was supported at JHU by the NSF under CAREER Award No.~DMR-0134377, 
at MIT by the NSF under contract 
number DMR-0071256, and at Toronto by the National 
Science and Engineering Research Council of Canada.  The work was also
supported by the NSF under Agreement No. DMR-9986442.  We 
acknowledge the support of NIST, U. S. Dept.~of Commerce, in 
providing the neutron facilities used in this work.


\end{document}